\documentclass[conference]{IEEEtran}
\IEEEoverridecommandlockouts

\usepackage[utf8]{inputenc}
\usepackage{amsmath,amssymb,amsthm}
\usepackage{accents,latexsym,cancel}
\usepackage{cite}

\usepackage[dvipsnames]{xcolor}
\definecolor{myPink}{RGB}{255,105,183}

\usepackage[T1]{fontenc}
\usepackage{graphics} 
\usepackage{epsfig} 
\usepackage[mathscr]{euscript}
\usepackage{algorithm}
\usepackage[noend]{algpseudocode}
\usepackage{comment}
\usepackage{bbm}
\makeatletter
\def\BState{\State\hskip-\ALG@thistlm}
\makeatother

\usepackage{tikz}
\usetikzlibrary{arrows,shapes,chains,matrix,positioning,scopes,patterns,calc,
decorations.markings,
decorations.pathmorphing,
}

\usepackage{pgfplots}
\pgfplotsset{compat=1.3}
\usepgflibrary{shapes}

\renewcommand{\epsilon}{\varepsilon}

\newcommand{\RNum}[1]{\uppercase\expandafter{\romannumeral #1\relax}}

\newcommand{\av}{\ensuremath{\mathbf{a}}}

\newcommand{\hv}{\ensuremath{\mathbf{h}}}
\newcommand{\mv}{\ensuremath{\mathbf{m}}}

\newcommand{\rv}{\ensuremath{\mathbf{r}}}
\newcommand{\sv}{\ensuremath{\mathbf{s}}}
\newcommand{\uv}{\ensuremath{\mathbf{u}}}
\newcommand{\vv}{\ensuremath{\mathbf{v}}}

\newcommand{\xv}{\ensuremath{\mathbf{x}}}
\newcommand{\yv}{\ensuremath{\mathbf{y}}}
\newcommand{\zv}{\ensuremath{\mathbf{z}}}

\newcommand{\Am}{\ensuremath{\mathbf{A}}}
\newcommand{\Bm}{\ensuremath{\mathbf{B}}}

\newcommand{\Hm}{\ensuremath{\mathbf{H}}}
\newcommand{\Idm}{\ensuremath{\mathbf{I}}}
\newcommand{\Sm}{\ensuremath{\mathbf{S}}}

\newcommand{\Wm}{\ensuremath{\mathbf{W}}}

\def\Pr{\mathrm{Pr}}

\DeclareMathAlphabet{\mcl}{OMS}{cmsy}{m}{n}

\newlength\tikzwidth
\newlength\tikzheight

\definecolor{mycolor1}{rgb}{0.63529,0.07843,0.18431}%
\definecolor{mycolor2}{rgb}{0.00000,0.44706,0.74118}%
\definecolor{mycolor3}{rgb}{0.00000,0.49804,0.00000}%
\definecolor{mycolor4}{rgb}{0.87059,0.49020,0.00000}%
\definecolor{mycolor5}{rgb}{0.00000,0.44700,0.74100}%
\definecolor{mycolor6}{rgb}{0.74902,0.00000,0.74902}%

\title{HashBeam: Enabling Feedback Through Downlink Beamforming in Unsourced Random Access}
\author{Jamison R. Ebert, Krishna R. Narayanan, Jean-Francois Chamberland\\
Department of Electrical and Computer Engineering, Texas A\&M University
\thanks{
This material is based upon work supported, in part, by the National Science Foundation (NSF) under Grant CCF-2131106.}
}

\begin{document}

\maketitle

\begin{abstract}
Unsourced random access (URA) has emerged as a candidate paradigm for massive machine-type communication (MTC) in next-generation wireless networks. 
While many excellent uplink schemes have been developed for URA, these schemes do not specify a mechanism for providing feedback regarding whether a user's message was successfully decoded.
While this may be acceptable in some MTC scenarios, the lack of feedback is inadequate for applications that demand a high level of reliability. 
However, the problem of providing feedback to active users is complicated by the fact that the base station does not know the identities of the active users. 
In this paper, a novel downlink beamforming scheme called \textit{HashBeam} is presented that enables the base station to provide feedback to the active users within URA, despite not knowing their identities. 
The key idea of this scheme is that the users' channels and hashes of their messages may be used as proxies for their true identities. 
The proposed scheme may be adapted to any number of antennas at the base station and it is shown that the required number of channel uses is linear in the number of users to acknowledge. 
The idea of using channel gains in conjunction with user hashes as discriminating attributes of active users is novel and expands the design space of URA schemes. 

\end{abstract}


\section{Introduction}

A significant challenge in next-generation wireless networks is that of supporting a very large number of unattended, machine-type devices. 
In addition to dramatically increasing network densities, these devices pose a unique challenge because they utilize network resources in a fundamentally different way compared to traditional human-operated devices.
Indeed, these devices are envisioned to sporadically transmit very short messages as opposed to sustaining long connections and transmitting large amounts of data.
Existing network access protocols become very inefficient under such traffic; thus, novel physical and data-link layer processes must be developed to accommodate these users.

One popular solution to this challenge of machine-type communication (MTC) is the paradigm of unsourced random access (URA)~\cite{polyanskiy2017perspective,vem2017user}. 
URA is able to support an arbitrarily large number of connected users as long as only a small subset of those users is active at any given point in time. 
This capability is enabled by the fact that there is no fine multi-user coordination under URA; rather, each device employs the exact same codebook when communicating with the central base station. 
At the receiver, a list of transmitted messages is recovered without regard to the identities of the senders. 
This strategy allows the network to avoid the overhead associated with user scheduling, which is especially important in MTC wherein this cost cannot be amortized over long payloads. 

Over the past several years, significant research has been dedicated to finding low-complexity schemes for the URA \textit{uplink} channel when the base station is equipped with a single antenna (see, e.g., \cite{calderbank2018chirrup, amalladinne2019coded, fengler2019sparcs, pradhan2019polar, amalladinne2020unsourced, ebert2021codeddemixing, han2021sparse, ahmadi2021random, nassaji2022unsourced, andreev2020polar}) and when the base station is equipped with an array of antennas (see \cite{shyianov2020massive, maxime2021tensor, fengler2021mimo, fengler2022pilotbased, gkagkos2022fasura}).
However, all of these schemes operate under the same premise: each active user has no way of knowing whether its message was correctly recovered.
If the base station fails to decode a message, that message is lost forever. 
While such behavior is tolerable for many MTC applications, this situation is unacceptable for applications that demand a higher level of reliability.  
This points to the need for an adequate feedback mechanism tailored to URA systems.

Independent of URA, hybrid automatic repeat request (HARQ) and its variants have found considerable success as pragmatic means for providing active users with enhanced connectivity.
This class of access schemes relies on timely feedback from the base station regarding whether messages have been correctly recovered.
To implement such a scheme under URA, one possibility is for the base station to send an acknowledgement (ACK) to every user whose message was successfully decoded. 
Therein, when an active user receives an ACK, it is finished with its transmission.
Conversely, when an active user does not receive an ACK, it transmits additional parity symbols to the base station, which then attempts to decode the message by leveraging the original transmission in addition to the extra parity bits.
Optionally, this feedback-retransmission process may repeat itself several times. 
This type of strategy is very effective and has been widely adopted as HARQ and its variants play a crucial role in many modern wireless networking standards.
We view the ability to provide timely feedback as a key step in implementing HARQ in URA settings.

The application of HARQ to URA has only recently been considered.
In \cite{popovski2022}, Kal{\o}r et al.\ seek to acknowledge $K$ users by broadcasting a common feedback signal to all $N$ users in the network. 
Though their scheme assumes that every user possesses a unique identifier, their scheme may be adapted to URA by using a portion of the user's message as the identifier. 
They point out that, naively, one may simply transmit the concatenation of the $K$ identifiers; however, this requires the transmission of $K \log_2 (N)$ bits, which seems unacceptably large for typical values of $N$.
The required number of bits may be reduced by enumerating all $\binom{N}{K}$ subsets and transmitting the index of the subset corresponding to the current set of decoded users; this requires only $\left\lceil \log_2 \binom{N}{K} \right\rceil$ bits.  
A principal result of \cite{popovski2022} is that, if a small amount of false positives are allowed, the number of bits can be further reduced to $K \lceil \log_2 \left( 1/\epsilon_{\mathrm{fp}} \right) \rceil$, where $\epsilon_{\mathrm{fp}}$ is the rate of false positives introduced by the scheme, by solving a set of linear equations over a Galois field. 

In \cite{yang2016csack}, the challenge of providing feedback is considered from a compressed sensing (CS) perspective.
Therein, a $K$-sparse vector of length $N$ is constructed where the non-zero entries indicate the $K$ out of $N$ users to be acknowledged. 
This vector is compressed and transmitted across the channel, and then the active users employ regularized approximate message passing to reconstruct the sent vector and determine whether their message was correctly decoded. 
To apply this scheme to URA, the users' messages would have to be used as proxies for their identifiers. 
Though elegant, this scheme requires many more channel uses than the scheme proposed in \cite{popovski2022}.

As a side remark, we note that the problem of providing minimal length feedback to active users was also considered by Kang and Yu in \cite{yu2021minimumfeedback}; yet, this work pertains to collision-limited scheduling of active users. 
Though one could think of providing feedback in terms of scheduling the active users into two bins based on whether their messages were successfully decoded, the results of \cite{yu2021minimumfeedback} cannot be directly applied to the URA problem because it relies on random binning.
A key distinction stems from the fact that in scheduling, it does not matter which users are assigned to which bins, as long as the total number of users in a single bin does not exceed a threshold.
This is obviously not the case for feedback.

Existing results rely on broadcasting a common message to all devices in the network. 
Under the paradigm of sending a common message as feedback, the proposed solutions are indeed highly effective.
However, each user is ultimately only concerned about whether its own message was decoded successfully. 
Furthermore, the base station is often equipped with multiple antennas and, in such cases, the URA feedback scheme should exploit the additional degrees of freedom afforded by the antenna array.

\subsection{Main Contribution}

In this article, a novel downlink beamforming scheme for the URA broadcast channel is presented.
This paradigm, which we call \emph{HashBeam}, uses a combination of beamforming and hashing to provide \textit{individualized} binary feedback to every active user. 
Crucial to this scheme is the insight that, while the base station does not know the identities of the active users, it does possess channel estimates along with the content of decoded uplink messages.
A combination of the channel gains and a hash of the recovered data can therefore be leveraged to construct an effective feedback mechanism, without explicitly using the identities of the senders.
Specifically, the users' channels are first exploited to send feedback \textit{in the direction of the origin of each} message. 
Then, because multiple devices may have nearby channel realizations in the signal space, the users' hashes are exploited to further separate sources.
This ensures that pertinent feedback can be delivered directly to every transmitter, albeit in an anonymous fashion.
The scheme presented in this paper leverages hashes and channel gains within the beamforming step and may be adapted to any number of antennas at the base station by adjusting the required length of the hash. 
In Section~\ref{sec:system_performance}, it is shown that the required number of channel uses grows linearly with the number of users to acknowledge, thus matching the order-wise performance of the scheme presented in \cite{popovski2022} without sending a common message to all users in the network. 

\begin{figure}
    \centering
    \begin{tikzpicture}
  [
  font=\footnotesize, draw=black, >=stealth', line width=1.25pt,
  base_station/.style={rectangle, minimum height=20mm, minimum width=15mm, draw=black, rounded corners},
  user_equipment_decoded/.style={rectangle, minimum height=7mm, minimum width=5mm, draw=blue, rounded corners},
  user_equipment_not_decoded/.style={rectangle, minimum height=7mm, minimum width=5mm, draw=red, dashed, rounded corners}
  ]

\node[base_station, align=center] (base_station_1) at (0, 0) {Base \\ Station};

\foreach \x in {0, 1, 2, 4} {
    \draw[line width=1pt, color=black, -triangle 90] plot[smooth, tension=0] coordinates {(0.75, 0.8-\x*0.4)(1.25, 0.8-\x*0.4)(1.25, 1.1-\x*0.4)};
}

\node[user_equipment_decoded, align=center] (ue_1) at (4.5, 2) {UE};
\node[user_equipment_not_decoded, align=center] (ue_2) at (5.0, 0.95) {UE};
\node[user_equipment_not_decoded, align=center] (ue_3) at (4.0, -0.10) {UE};
\node[user_equipment_decoded, align=center] (ue_4) at (4.9, -1.25) {UE};

\draw[rotate=60] (1.52, -1.0) ellipse(0.25 and 0.05);
\draw[rotate=40] (2.25, -0.75) ellipse(1 and 0.15);
\draw (1.75, 0.1) ellipse(0.25 and 0.10);
\draw[rotate=-20] (2.40, 0.40) ellipse(1 and 0.15);
\draw[rotate=-45] (1.60, 0.85) ellipse(0.25 and 0.05);

\node[draw=none] (ue_1_hash) at (5.6, 2) {$\begin{bmatrix} \vrule \\ \av_i \\ \vrule \end{bmatrix}$};
\node[draw=none] (ue_2_hash) at (6.1, 0.95) {$\begin{bmatrix} \vrule \\ \av_j \\ \vrule \end{bmatrix}$};
\node[draw=none] (ue_3_hash) at (5.1, -0.1) {$\begin{bmatrix} \vrule \\ \av_k \\ \vrule \end{bmatrix}$};
\node[draw=none] (ue_4_hash) at (6.0, -1.25) {$\begin{bmatrix} \vrule \\ \av_l \\ \vrule \end{bmatrix}$};

\draw[<->, dashed, line width=1.0] (4.8, 2) -- (5.4, 2);
\draw[<->, dashed, line width=1.0] (5.3, 0.95) -- (5.9, 0.95);
\draw[<->, dashed, line width=1.0] (4.3, -0.1) -- (4.9, -0.1);
\draw[<->, dashed, line width=1.0] (5.2, -1.25) -- (5.8, -1.25);

\end{tikzpicture}
    \caption{HashBeam is a novel downlink beamforming scheme for the URA broadcast channel that leverages decoded users' channels in conjunction with hashes of their uplink messages to beamform acknowledgements to decoded users, despite not knowing those users' identities. In this figure, the user equipments (UE)s with solid lines are the ones to be acknowledged and $\av_i$ represents the hash of the $i$th users' data. }
    \label{fig:generic_system_model}
\end{figure}
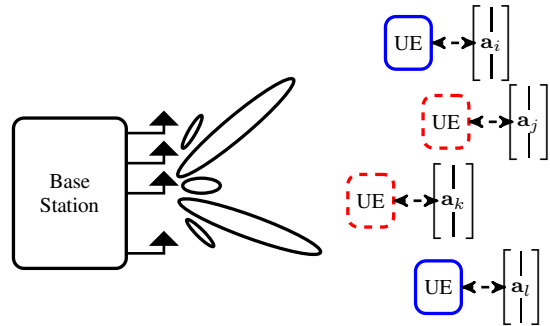

\subsection{Notation}
Matrices are denoted by capital letters such as $\Am$, and column vectors are denoted by lower-case bold letters such as $\xv$. 
For vectors $\uv$ and $\vv$, the standard inner product is denoted as $\langle \uv, \vv \rangle$. 
The Hermitian of $\Am$ is denoted by $\Am^{\mathrm{H}}$ and the Khatri-Rao product of $\Am$ and $\Bm$ is denoted as $\Am \ast \Bm$.
For vectors $\uv$ and $\vv$, the Kronecker product of $\uv$ and $\vv$ is denoted as $\uv \otimes \vv$.
For any positive integer $N$, $[N] = \{1, 2, \ldots, N\} \subset \mathbb{Z}_{+}$. 
Finally, $\Pr\left(\mathcal{X}\right)$ refers to the probability of event $\mathcal{X}$ and $\mathbb{E}[X]$ refers to the expected value of random variable $X$. 

\section{System Model}
\label{sec:system_model}

Consider a URA scenario in which there are $N$ users, out of which $K_a$, $K_a \ll N$, users are active at a given time instant.
Each user is equipped with a single antenna and the central base station is equipped with $M$ antennas. 
For simplicity of notation, each active user is assigned a unique but arbitrary label $i \in [K_a]$.
We emphasize that these labels are for notational purposes only and do not reveal any information about the true identities of these users. 
Within a frame, all active users simultaneously transmit their short uplink messages to the central base station according to some uplink URA scheme (see, e.g., \cite{han2021sparse, gkagkos2022fasura}).
In this article, the specifics of the URA scheme are unimportant and can therefore be abstracted away.
The focus is on the ability to provide feedback using downlink transmissions.
Still, we assume that a portion of the users are decoded correctly and we let $\mathcal{D} \subseteq [K_a]$ denote this set of successfully decoded users, where $|\mathcal{D}| = K$.
Without loss of generality, we assume that $\mathcal{D} = [K]$ throughout this paper. 
Furthermore, we assume that in the process of message decoding or as a result of this process, channel estimates for these same users are available at the base station.
Moreover, to keep the discussion simple, we assume that the channel estimates obtained by the base station are perfect. 
Note that once a message is decoded correctly, a hash can be computed based on this recovered message.

The channels between devices and the base station are independent quasi-static Rayleigh fading channels with coherence times at least as long as the duration of the communication cycle.
This implies that devices remain stationary for the duration of the transmission-feedback process and that channel reciprocity holds.
While the assumption of device stationarity is not always valid, it does hold for many classes of machine-type devices, especially when the end-to-end system latency is very small. 
We denote the channel between user~$i$ and the base station by $\hv_i \in \mathbb{C}^M$, where $\left( \hv_i \right)_j \sim \mathcal{CN} \left( 0, 1 \right)$. 

\subsection{Proposed Scheme}

Though the base station does not know the identities of the decoded users, it does have channel estimates and it can compute hashes associated with every successfully recovered message.
Under the aforementioned assumptions of stationarity and channel coherence, the base station can use the channels of the active users as proxies for their identities and provide message-specific feedback by beamforming ACKs \textit{in the direction of the origin} of each recovered message. 
This problem is closely related to that of downlink beamforming, which has been well-studied over the past decades. 
However, the task of providing feedback through downlink beamforming is exacerbated by the fact that the users' channels are often not statistically well-separated, especially when $M$ is less than $K$.
To handle this challenge, the hashes of the users' data are used in conjunction with their channels to create downlink beamforming vectors.
Thus, the base station is able to provide individualized feedback to successfully decoded users without explicitly knowing their true identities. 

Specifically, let the base station and all active devices have access to a hash function $f: \{0, 1\}^B \rightarrow \mathbb{C}^{L}$, which computes a length $L$ random hash based on $B$ bits of a user's message $\mv$. 
Let the $i$th user's hash be denoted by $\av_i$ and let $\Am$ denote the concatenation of all $K$ hashes:
\begin{equation}
    \Am = \begin{bmatrix}
    \vrule & \vrule & & \vrule \\
    \av_1 & \av_2 & \ldots & \av_K \\
    \vrule & \vrule & & \vrule 
    \end{bmatrix} \in \mathbb{C}^{L \times K} .
\end{equation}
Throughout this paper, we assume that the entries of $\Am$ are of the form $\Am_{i, k} = \alpha\exp\{j\phi_{i,k}\}$ and $\phi_{i,k} \sim \operatorname{Uniform}[0, 2\pi)$.  
Furthermore, let $\Hm$ denote the concatenation of all $K$ users' channels,
\begin{equation}
    \Hm = \begin{bmatrix}
    \vrule & \vrule & & \vrule \\
    \hv_1 & \hv_2 & \ldots & \hv_K \\
    \vrule & \vrule & & \vrule 
    \end{bmatrix} \in \mathbb{C}^{M \times K} .
\end{equation}
Throughout this article, we assume that the channel estimates are exact.
Yet, the techniques we present below can be extended to the more general setting by utilizing estimate $\hat{\Hm}$ instead of $\Hm$. 

In our proposed scheme, the base station begins by taking the Khatri-Rao (column-wise Kronecker) product of $\Am$ and $\Hm$ to obtain 
\begin{equation}
    \Sm = \Am \ast \Hm \in \mathbb{C}^{LM \times K}.
\end{equation}
Eventually, the base station employs a beamforming vector $\mathbf{w}_i$ to acknowledgement recovery of message~$i \in \mathcal{D}$.
Thus, the signal transmitted from the base station is of the form
\begin{equation} \label{equation:DownlinkInput}
    \mathbf{v} = \Wm \xv_d
\end{equation}
where $\xv_d^T = [ 1 \ 1 \ \cdots \ 1 ] \in \mathbb{R}^K$ and $\Wm$ is a $LM \times K$ matrix whose $i$th column is given by $\mathbf{w}_i$. 
The matrix $\mathbf{W}$ is designed using uplink downlink duality as explained below.

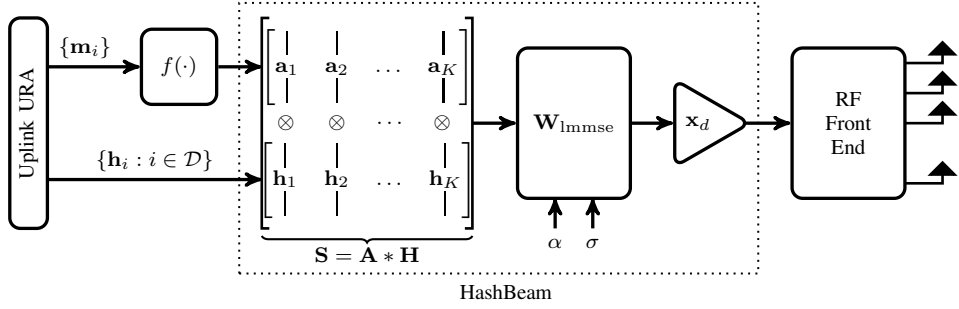
\begin{figure*}[t]
    \centering
    \begin{tikzpicture}
  [
  font=\footnotesize, draw=black, >=stealth', line width=1.25pt,
  uplink_ura/.style={rectangle, minimum height=5mm, minimum width=28mm, draw=black, rounded corners},
  black_box/.style={rectangle, minimum height=20mm, minimum width=15mm, draw=black, rounded corners},
  hash_function/.style={rectangle, minimum height=10mm, minimum width=10mm, draw=black, rounded corners},
  multiplication_triangle/.style ={isosceles triangle,isosceles triangle apex angle=60, minimum height=10mm, rounded corners, draw=black, fill=white}
  ]
 
\node[uplink_ura, rotate=90] (uplink_ura) at (-4.5, 0.0) {Uplink URA};

\node[hash_function] (hash_function_1) at (-2.5, 0.75) {$f(\cdot)$};
\draw[->] (-4.25, 0.75) -- (hash_function_1.west);
\draw[->] (hash_function_1.east) -- (-1.4, 0.75);
\draw[->] (-4.25, -0.75) -- (-1.4, -0.75);
\node[draw=none] at (-3.75, 1.0) {$\{\mv_i\}$};
\node[draw=none] at (-2.825, -0.5) {$\{\hv_i : i \in \mathcal{D}\}$};

\node[draw=none] at (0, 0.75) {$\begin{bmatrix}
    \vrule & \vrule & & \vrule \\
    \av_1 & \av_2 & \ldots & \av_K \\
    \vrule & \vrule & & \vrule 
    \end{bmatrix}$};
    
\node[draw=none] at (0, -0.75) {$\begin{bmatrix}
    \vrule & \vrule & & \vrule \\
    \hv_1 & \hv_2 & \ldots & \hv_K \\
    \vrule & \vrule & & \vrule 
    \end{bmatrix}$};
    
\node[draw=none] at (-1.08, 0.0) {$\otimes$};
\node[draw=none] at (-0.4, 0.0) {$\otimes$};
\node[draw=none] at (0.3, 0.0) {$\ldots$};
\node[draw=none] at (1.0, 0.0) {$\otimes$};

\draw[line width=1.0] (-1.3, -1.4) -- (-1.4, -1.4) -- (-1.4, 1.4) -- (-1.3, 1.4);
\draw[line width=1.0] (1.3, -1.4) -- (1.4, -1.4) -- (1.4, 1.4) -- (1.3, 1.4);
\draw[thick, decoration={brace, mirror}, decorate] (-1.4, -1.5) -- (1.4, -1.5);
\node[draw=none] at (0.0, -1.75) {$\Sm = \Am \ast \Hm$};

\node[black_box, align=center] (w_lmmse) at (2.75, 0) {$\Wm_{\mathrm{lmmse}}$};
\draw[->] (1.4, 0.0) -- (w_lmmse.west);
\draw[->] (2.5, -1.4) -- (2.5, -1.0);
\node[draw=none] at (2.5, -1.6) {$\alpha$};
\draw[->] (3.0, -1.4) -- (3.0, -1.0);
\node[draw=none] at (3.0, -1.6) {$\sigma$};

\node[multiplication_triangle, align=center] (x_mult) at (4.4, 0) {$\xv_d$};
\draw[->] (w_lmmse.east) -- (x_mult.west);

\draw[->, dotted, line width=0.75] (-1.7, -2.0) rectangle (5.2, 1.6);
\node[draw=none] at (1.85, -2.25) {HashBeam};

\node[black_box, align=center] (rf_chain) at (6.4, 0) {RF \\ Front \\ End};
\draw[->] (5.0, 0.0) -- (rf_chain.west);

\draw [line width=1pt,color=black,-triangle 90] plot[smooth, tension=0] coordinates {(7.15,0.8)(7.65,0.8)(7.65,1.1)};
\draw [line width=1pt,color=black,-triangle 90] plot[smooth, tension=0] coordinates {(7.15,0.4)(7.65,0.4)(7.65,0.7)};
\draw [line width=1pt,color=black,-triangle 90] plot[smooth, tension=0] coordinates {(7.15,0)(7.65,0)(7.65,0.3)};
\draw [line width=1pt,color=black,-triangle 90] plot[smooth, tension=0] coordinates {(7.15,-0.8)(7.65,-0.8)(7.65,-0.5)};
  
\end{tikzpicture}
    \caption{This figure illustrates the downlink beamforming process employed by HashBeam. 
    The uplink URA scheme provides channel estimates and hashes for the uplink message associated with each decoded user. 
    The Khatri-Rao product of the hashes and the channels is computed, and downlink beamforming vectors are computed via uplink-downlink duality. 
    The sum of all beamforming vectors is then transmitted by the base station. }
    \label{fig:beamforming_diagram}
\end{figure*}

The dual uplink channel is modelled as 
\begin{equation}
    \yv = \Sm \xv_u + \zv_u \in \mathbb{C}^{LM},
\end{equation}
where $\xv_u$ represents the aggregate uplink input and $\zv_u$ is a vector of circularly-symmetric additive white Gaussian noise (AWGN) with zero-mean and covariance $\alpha^2\sigma^2\Idm$. 
For this uplink dual, a good estimate for $\xv_u$ given $\yv$ in the mean-square error sense is given by the linear minimum mean square error (LMMSE) estimate $\hat{\xv}_{u} = \Wm_{\mathrm{lmmse}}^\mathrm{H}\yv$, where 
\begin{equation}
    \label{eq:lmmsebeamforming}
    \Wm_{\mathrm{lmmse}}^\mathrm{H} = \left(\alpha^2\sigma^2\Idm + \Sm^\mathrm{H}\Sm\right)^{-1}\Sm^\mathrm{H} \in \mathbb{C}^{K \times LM}.
\end{equation}
Under the proper power allocation scheme, column $i$ of $\Wm_{\mathrm{lmmse}}$ is the optimal downlink beamforming vector for user $i$ in a traditional communication scenario.
Inspired by this result, we set $\Wm = \Wm_{\mathrm{lmmse}}$.
Thus, \eqref{equation:DownlinkInput} becomes
\begin{equation}
    \label{eq:finalbeamformingvector}
    \begin{split}
        \vv &= \Wm_{\mathrm{lmmse}}\xv_d \in \mathbb{C}^{LM} \\
    \end{split}
\end{equation}
and this signal is transmitted from the base station to all of the users using $M$ antennas over $L$ channel uses. 
Fig.~\ref{fig:beamforming_diagram} graphically depicts this process. 

The signal received by user $i$ is given by
\begin{equation}
    \rv_i = \begin{bmatrix}
    \langle \hv_i, \vv_1 \rangle + \zv_{i, 1}, & \ldots, & \langle \hv_i, \vv_{L} \rangle + \zv_{i, L}
    \end{bmatrix} \in \mathbb{C}^{L},
\end{equation}
where $\vv_j \triangleq \vv\left((j-1)M+1:jM\right)$ and $\zv_{i, j} \sim \mathcal{CN}\left(0, \sigma^2\right)$ is the $j$th element of the $i$th user's noise vector. 
User $i$ will then correlate its received signal with its hash $\av_i$ to create statistic
\begin{equation}
    \theta_i = \langle \av_i, \rv_i \rangle \in \mathbb{C}.
\end{equation}
The receiver then performs a binary hypothesis test to determine whether an ACK was received. 
When $\theta_i \in \mathcal{R}_0$, the user fails to reject the null hypothesis $\mathcal{H}_0$ and assumes that no ACK has been received. 
Conversely, if $\theta_i \in \mathcal{R}_1$, the user rejects the null hypothesis in favor of $\mathcal{H}_1$ and decodes an ACK. 
We note that, in this scheme, each active user can only determine whether its own message was successfully decoded; this is what is refer to as individualized feedback. 

\subsection{Proposed Hypothesis Test}
To develop a good hypothesis test, we must first understand the nature of the test statistic $\theta_i$. 
We begin by expressing $\theta_i$ as follows
\begin{equation}
    \begin{split}
        \theta_i &= \langle \av_i, \rv_i \rangle \\
        &= \sum_{j = 1}^L \left( \av_{i, j}^* \langle \hv_i, \vv_j \rangle + \av_{i,j}^* \zv_{i,j} \right) \\
        &= \sum_{j = 1}^L \left( \langle \av_{i, j}\hv_i, \vv_j \rangle + \av_{i,j}^* \zv_{i,j} \right) \\
        &= \langle \sv_i, \vv \rangle + \langle \av_i, \zv_i \rangle. \\
    \end{split}
    \label{eqn:thetai}
\end{equation}
To get some insight into the distribution of $\theta_i$, we first consider the case when $\Sm^{\mathrm{H}} \Sm$ is diagonal with the diagonal entries given by $\| \sv_j \|_2^2, j \in \mathcal{D}$. 
Furthermore, we also neglect the $\alpha^2\sigma^2\Idm$ term within the computation of $\Wm_{\mathrm{lmmse}}$. 
In this case, it can be seen that the first term in \eqref{eqn:thetai} reduces to $\sum_{m \in \mathcal{D}} \frac{\langle \sv_{i},\sv_m \rangle}
{\langle \sv_m, \sv_m \rangle}$. 
Hence, under the aforementioned simplifications, 
\begin{equation}
    \theta_{i} = \begin{cases}
    1 + \langle \av_i, \zv_i \rangle, & i \in \mathcal{D} \\
    \sum_{m \in \mathcal{D}} 
\frac{\langle \sv_{i},\sv_m \rangle}
{\langle \sv_m, \sv_m \rangle} + \langle \av_i, \zv_i \rangle, & i \in [K_a] \setminus \mathcal{D}.
\end{cases}
\end{equation}
Thus, for the correctly decoded users, i.e., for $i \in \mathcal{D}$, $\theta_i \sim \mathcal{CN}(1, L\alpha^2\sigma^2)$ whereas, for the users who are not correctly decoded, i.e. $i \in [K_a] \setminus \mathcal{D}$, $\theta_i$ is a zero-mean complex random variable. 

Since the number of decoded messages, their hashes, and the corresponding channel gains are random; it is not necessarily the case that $\langle \sv_i, \sv_j \rangle = 0$.
Thus, in reality, $\theta_i$ is a random variable whose distribution may not be complex Gaussian.
The model intricacies associated with the distribution of $\theta_i$ make finding the optimal decision region quite challenging.
In this article, we circumvent this issue by approximating $\theta_i$ as complex Gaussian for both successfully and unsuccessfully decoded users. 
Let $\theta_i$ have mean $\mu_0$ and variance $\sigma^2_0$ when $i \in [K_a] \setminus \mathcal{D}$ and mean $\mu_1$ and variance $\sigma^2_1$ when $i \in \mathcal{D}$. 
As normally $\sigma^2_0 \neq \sigma^2_1$, a quadratic discriminant may be computed using a Neyman-Pearson test to separate $\mathcal{R}_0$ and $\mathcal{R}_1$. 
The distribution parameters $\mu_0$, $\sigma^2_0$, $\mu_1$, and $\sigma^2_1$ can be approximated through sampling.


\section{System Performance}
\label{sec:system_performance}

The predominant performance metrics for this downlink beamforming scheme are the probability of false alarm $P_{\mathrm{FA}}$ and the probability of a miss $P_{\mathrm{MD}}$. 
A false alarm occurs when user $i$'s uplink message is not successfully decoded (i.e. $i \in [K_a] \setminus \mathcal{D}$) yet user $i$ decodes an ACK because $\theta_i \in \mathcal{R}_1$.
Conversely, a miss occurs when user $i$'s message is successfully decoded (i.e. $i \in \mathcal{D}$) but user $i$ does not declare an ACK because $\theta_i \in \mathcal{R}_0$
Within the context of HARQ, the cost of a miss is relatively low as it only results in an unnecessary transmission of additional parity symbols. 
False alarms, on the other hand, are much more costly as they result in a user's message being lost forever. 
Herein, we are interested in the regime where both $P_{\mathrm{FA}}$ and $P_{\mathrm{MD}}$ are less than $0.05$.

For a fixed number of antennas $M$, an interesting question to ask is how the hash length $L$ must scale as a function of the number of successfully decoded users $K$. 
Clearly, we would like to minimize $L$ as this quantity is equal to the required number of channel uses for downlink beamforming. 
We begin by analyzing the noiseless case, i.e., $\sigma^2 = 0$. 
A straightforward lower bound for the proposed scheme in the noiseless setting is $L \geq \frac{K}{M}$; this bound ensures that the matrix product $\Sm^\mathrm{H}\Sm$ in \eqref{eq:lmmsebeamforming} has full rank (almost surely) and is therefore invertible. 
Numeric simulations shows that the proposed scheme achieves this lower bound for all values of $K$ and $M$ considered; thus $L = \lceil\frac{K}{M}\rceil$ suffices in the noiseless case. 
We note that this lower bound is inherent to the proposed scheme and not to the general problem of downlink beamforming. 

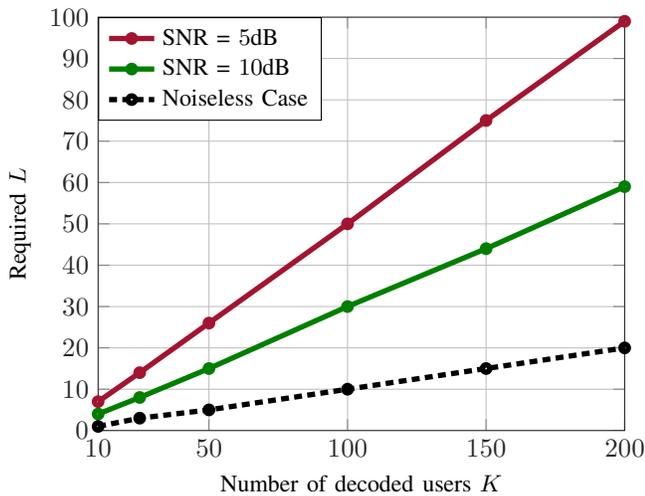
\begin{figure}[t]
    \centering
    \begin{tikzpicture}

\definecolor{customred}{rgb}{0.63529,0.07843,0.18431} 
\definecolor{customblue}{rgb}{0.00000,0.44706,0.74118} 
\definecolor{customgreen}{rgb}{0.00000,0.49804,0.00000} 

\begin{axis}[
    font=\small,
    width=7cm,
    height=5.5cm,
    scale only axis,
    every outer x axis line/.append style={white!15!black},
    every x tick label/.append style={font=\color{white!15!black}},
    xmin=10,
    xmax=200,
    xtick = {10, 50, 100, 150, 200},
    xlabel={Number of decoded users $K$},
    xmajorgrids,
    every outer y axis line/.append style={white!15!black},
    every y tick label/.append style={font=\color{white!15!black}},
    ymin=0,
    ymax=100,
    ytick = {0, 10, 20, ..., 100},
    ylabel={Required $L$},
    ymajorgrids,
    legend style={at={(0,1)},anchor=north west, draw=black,fill=white,legend cell align=left}
]

\addplot [
    color=customred,
    solid,
    line width=2.0pt,
    mark size=1.4pt,
    mark=o,
    mark options={solid}
]
table[row sep=crcr]{
    10 7 \\
    25 14 \\
    50 26 \\
    100 50 \\
    150 75 \\
    200 99 \\
};
\addlegendentry{SNR = $5$dB};

\addplot [
    color=customgreen,
    solid,
    line width=2.0pt,
    mark size=1.4pt,
    mark=o,
    mark options={solid}
]
table[row sep=crcr]{
    10 4 \\    %
    25 8 \\    
    50 15 \\   
    100 30 \\  
    150 44 \\  
    200 59 \\  
};
\addlegendentry{SNR = $10$dB};

\addplot [
    color=black,
    densely dashed,
    line width=2.0pt,
    mark size=1.4pt,
    mark=o,
    mark options={solid}
]
table[row sep=crcr]{
    10 1 \\
    25 3 \\
    50 5 \\
    100 10 \\
    150 15 \\
    200 20 \\
};
\addlegendentry{Noiseless Case};

\end{axis}
\end{tikzpicture}
    \caption{Required hash length $L$ to obtain a $P_{\mathrm{MD}} \leq 0.05$ and $P_{\mathrm{FA}} \leq 0.05$ as a function of the number of successfully decoded users. Here, results are presented for the case of $M = 10$ antennas at base station.}
    \label{fig:LvsKfixedMSNR}
\end{figure}

We also consider the performance of this scheme in the presence of AWGN. 
In such circumstances, we define the signal to noise ratio ($\operatorname{SNR}$) to be
\begin{equation}
    \label{eq:snr}
    \begin{split}
        \operatorname{SNR} &= \frac{\mathbb{E}\left[ \sum_{j=1}^L |\langle \hv, \vv_j \rangle|^2\right]}{\mathbb{E}\left[\|\zv\|_2^2\right]}
        = \frac{\mathbb{E}\left[ \sum_{j=1}^L |\langle \hv, \vv_j \rangle|^2\right]}{L\sigma^2} .
    \end{split}
\end{equation}
For fixed $\sigma^2$, one may obtain the desired $\operatorname{SNR}$ by adjusting $\alpha$, which is inversely proportional to $\operatorname{SNR}$.
Fig.~\ref{fig:LvsKfixedMSNR} plots the required hash length $L$ as a function of $K$ to achieve $P_{\mathrm{MD}} \leq 0.05$ and $P_{\mathrm{FA}} \leq 0.05$ when $M = 10$ for various SNRs. 
In addition, Fig.~\ref{fig:LvsSNR} shows the required $L$ as a function of $K$ to achieve the target error rates when $\operatorname{SNR} = 10$~dB for various numbers of antennas $M$. 
These results highlight that the proposed scheme may be easily adapted to any combination of number of antennas and $\operatorname{SNR}$ by simply adjusting the length of the hash.
From Fig.~\ref{fig:LvsKfixedMSNR}, Fig.~\ref{fig:LvsSNR}, and our noiseless results, it is also clear that the required hash length $L$, and by extension, the required number of channel uses, scales as $\mathcal{O}\left(K\right)$; thus our practical scheme scales on the same order as the theoretic results presented in \cite{popovski2022}. 
We emphasize that this scaling is independent of the total number of users $N$ and the number of active users $K_a$. 

\section{Conclusion}

In this article, a novel downlink beamforming scheme entitled \emph{HashBeam} is presented that allows the base station to provide individualized feedback to successfully decoded users, despite not knowing the identities of those users. 
To accomplish this task, the base station leverages knowledge of the decoded users' channels as well as hashes of their uplink messages to beamform feedback directly to the decoded users. 
The proposed scheme may be easily adapted to any number of antennas at the base station and to various $\operatorname{SNR}$s by simply adjusting the length of the hash, which scales as $\mathcal{O}\left(K\right)$. 

\begin{figure}
    \centering
    \begin{tikzpicture}

\definecolor{customred}{rgb}{0.63529,0.07843,0.18431} 
\definecolor{customblue}{rgb}{0.00000,0.44706,0.74118} 
\definecolor{customgreen}{rgb}{0.00000,0.49804,0.00000} 

\begin{axis}[
    font=\small,
    width=7cm,
    height=5.5cm,
    scale only axis,
    every outer x axis line/.append style={white!15!black},
    every x tick label/.append style={font=\color{white!15!black}},
    xmin=10,
    xmax=200,
    xtick = {10, 50, 100, 150, 200},
    xlabel={Number of decoded users $K$},
    xmajorgrids,
    every outer y axis line/.append style={white!15!black},
    every y tick label/.append style={font=\color{white!15!black}},
    ymin=0,
    ymax=60,
    ytick = {0, 10, 20, ..., 60},
    ylabel={Required $L$},
    ymajorgrids,
    legend style={at={(0,1)},anchor=north west, draw=black,fill=white,legend cell align=left}
]

\addplot [
    color=customred,
    solid,
    line width=2.0pt,
    mark size=1.4pt,
    mark=o,
    mark options={solid}
]
table[row sep=crcr]{
    10 4 \\    
    25 8 \\    
    50 15 \\   
    100 30 \\  
    150 44 \\  
    200 59 \\  
};
\addlegendentry{M = $10$};

\addplot [
    color=customgreen,
    solid,
    line width=2.0pt,
    mark size=1.4pt,
    mark=o,
    mark options={solid}
]
table[row sep=crcr]{
    10 3 \\     
    25 5 \\     
    50 8 \\     
    100 15 \\   %
    150 22 \\   
    200 29 \\   
};
\addlegendentry{M = $20$};

\addplot [
    color=customblue,
    solid,
    line width=2.0pt,
    mark size=1.4pt,
    mark=o,
    mark options={solid}
]
table[row sep=crcr]{
    10 2 \\
    25 3 \\
    50 4 \\
    100 7 \\
    150 9 \\
    200 12 \\
};
\addlegendentry{M = $50$};


\end{axis}
\end{tikzpicture}
    \caption{Required hash length $L$ to obtain a $P_{\mathrm{MD}} \leq 0.05$ and $P_{\mathrm{FA}} \leq 0.05$ as a function of the number of successfully decoded users. Here, results are presented for the case when $\operatorname{SNR} = 10$dB.}
    \label{fig:LvsSNR}
\end{figure}


\bibliographystyle{IEEEbib}
\bibliography{ITA2022.bib}

\end{document}